\documentstyle[nato,numreferences,epsf,graphicx,times,mathptm,cite]{crckapb}
\def\Olejnik{Olej\-n{\'\i}k, \v{S}.}
\begin{opening}
\title{Status of center dominance in various center gauges}
\author{Manfried Faber}
\institute{Atominstitut der \"osterreichischen Universit\"aten,\\
           Arbeitsgruppe Kernphysik, TU Wien, A--1040 Vienna, Austria}
\author{Jeff Greensite}
\institute{Physics and Astronomy Dept., San Francisco State University,\\
           San Francisco, CA 94117, USA, and \\
           Theory Group, Lawrence Berkeley National Laboratory,\\
           Berkeley, CA 94720, USA}
\author{\v{S}tefan Olejn{\'i}k}
\institute{Institute  of Physics, Slovak Academy of Sciences,\\
           SK--842 28 Bratislava, Slovakia}
\runningtitle{Status of Center Dominance}
\runningauthor{M.\ Faber, J.\ Greensite, \v{S}.\ Olejn{\'\i}k\ }
\end{opening}
\begin{document}
%
%
\begin{abstract}
We review arguments for center dominance in center gauges 
where vortex locations are correctly identified. We introduce
an appealing interpretation of the maximal center gauge, discuss
problems with Gribov copies, and a cure to the problems through 
the direct Laplacian center gauge. We study correlations
between direct and indirect Laplacian center gauges.
\end{abstract}
%
%
\renewcommand{\thefootnote}{\fnsymbol{footnote}}
\footnotetext[0]{Presented by {\v S}.\ Olejn{\'\i}k
at the NATO Advanced Research Workshop ``Confinement, Topology,
and other Non-Perturbative Aspects of QCD'',
January 21--27, 2002, Star\'a Lesn\'a (Slovakia).  
Supported in part by the NATO Collaborative Linkage Grant No.\ 
PST.CLG.976987 and by
the Slovak Grant Agency for Science, Grant No.\ 2/7119/2000.}
\renewcommand{\thefootnote}{\arabic{footnote}}
%
%
\vspace*{-0.5cm}
\section{{Why Center Dominance?}}
	The aim of most lattice studies of the confinement mechanism
is to extract from lattice link variables the most relevant parts
for the infrared dynamics. The concept of (some kind of)
dominance seems a necessary, though not sufficient, condition for success.
If we extract the (would-be) relevant parts of links and compute 
physical quantities related to confinement (e.g.\ the string tension)
we expect to reproduce their behavior in the full theory. Were it not the 
case, one could hardly claim to have achieved the goal.

	It was observed both in SU(2)~\cite{%
DelDebbio:1996mh,DelDebbio:1998uu} and (less convincingly) 
in SU(3) lattice gauge theory~\cite{Faber:1999sq} that the string tension
obtained from center-projected configurations in maximal center gauge (MCG) 
agrees remarkably well with the asymptotic string tension of the full 
theory.	This phenomenon of \textit{center dominance} has led to the recent 
revival of interest in the center-vortex picture of color 
confinement~\cite{tHooft:1977hy,Mack:1980rc,Proceedings:2002aa}. 
One can easily formulate an argument why center dominance 
\textit{should} occur if center vortices are correctly 
identified~\cite{DelDebbio:1998uu}: 

	Vortices are created by discontinuous gauge transformations.  
Let a closed loop~$C$, parametrized by $x^\mu(\tau)$, $\tau \in [0,1]$,
encircle $n$ vortices.  At the point of discontinuity (in SU(2)):
\begin{equation}
g(x(0)) = (-1)^n g(x(1)).
\end{equation}
The corresponding vector potential in the neighborhood of $C$ can
be decomposed as
\begin{equation}
A^{(n)}_\mu(x) = g^{-1}\delta A^{(n)}_\mu(x) g + 
i g^{-1} \partial_\mu g.
\end{equation}
The $g^{-1} \partial_\mu g$ term is dropped at the discontinuity.
Then, the value of the Wilson loop is
\begin{equation}
W_n(C) =\langle\mbox{Tr}\exp[i\oint dx^\mu A_\mu^{(n)}]\rangle 
=(-1)^n \langle\mbox{Tr}\exp[i\oint dx^\mu \delta A_\mu^{(n)}]\rangle.
\end{equation}
In the region of the loop $C$, the vortex background looks locally like
a gauge transformation.  If all other fluctuations $\delta A^{(n)}_\mu$ are
basically short-range, then they should be oblivious, in the neighborhood
of the loop $C$, to the presence or absence of vortices 
in the middle of the loop.  In that case:
\begin{equation}
\langle\mbox{Tr}\exp[i\oint dx^\mu \delta A_\mu^{(n)}\rangle
\approx \langle\mbox{Tr}\exp[i\oint dx^\mu \delta A_\mu^{(0)}]\rangle
\end{equation}
for sufficiently large loops, and therefore
\begin{equation}\label{Wn/W0}
{W_n(C)/W_0(C)} \longrightarrow (-1)^n
\quad\mbox{or}\quad
W(C)\approx Z(C)\times\langle
\mbox{Tr}\exp[i\oint dx^\mu \delta A_\mu^{(0)}]\rangle.
\end{equation}
Here $W(C)$ is the expectation value of the full Wilson loop, and $Z(C)$
the expectation value of the loop constructed from center elements alone.

	It is clear from the above argument that one gets center dominance,
i.e. the same string tension from $W(C)$ and $Z(C)$, under four intertwined 
assumptions:
\begin{enumerate}
\item
	Vortices are the confinement mechanism.
\item
	Vortices are correctly identified.
\item
	Short-range fluctuations with/without vortices look similar.
\item
	No area law arises from the last factor in Eq.\ (\ref{Wn/W0}).
\end{enumerate}

	Is there a necessity to fix any gauge? The original vortex idea was 
formulated without a reference to a particular gauge, and in fact a kind 
of center dominance exists even without gauge fixing, as was shown 
in \cite{Faber:1998en}. However, this holds for any distances, not only for
large ones, vortices defined without gauge fixing do not fulfill 
simple expectations and do not scale according to the renormalization group,
and thus the phenomenon hardly bears any information on the confinement
mechanism. Gauge fixing appears of special importance for correct
identification of vortices.

%
%
\section{{How to Identify Center Vortices?}}
	The procedure, proposed in~\cite{DelDebbio:1996mh,DelDebbio:1998uu}, 
consists of three steps:
\begin{enumerate}
\item
	Fix thermalized SU(2) lattice configurations to 
\textit{direct maximal center} (or adjoint Landau) 
\textit{gauge} by maximizing the expression:
\begin{equation}\label{MCG}
\sum_{x,\mu}\;\Bigl| \mbox{Tr}[U_\mu(x)] \Bigr|^2\;
\qquad\mbox{or equivalently}\qquad
       \sum_{x,\mu}\;\mbox{Tr}[U^A_\mu(x)]\;.
\end{equation}
\item
	Make \textit{center projection} by replacing:
\begin{equation}\label{projection}
      U_\mu(x)\rightarrow Z_\mu(x) \equiv \mbox{sign Tr}[U_\mu(x)]\;.
\end{equation}
\item
	Finally, identify excitations ({\em P-vortices\/}) of the resulting
$Z_2$ lattice configurations.
\end{enumerate}

	A whole series of results, obtained by our and other groups, 
indicates that center vortices defined in MCG play a crucial role in the
confinement mechanism. This includes, besides center dominance, the 
following:
\begin{enumerate}
\item
	P-vortices locate center vortices in full lattice 
configurations~\cite{DelDebbio:1998uu}. 
\item
	P-vortices locate physical objects, their density scales
according to the renormalization group~\cite{Langfeld:1997jx}.
\item
	Creutz ratios computed from center-projected
Wilson loops are almost constant starting from shortest distances; 
the Coulomb contribution was effectively eliminated
(\textit{precocious linearity})~\cite{DelDebbio:1998uu}.
\item
	Center vortices are correlated not only with 
confinement, but with chiral symmetry
breaking and non-trivial topology as well~\cite{deForcrand:1999ms}.
\item
	Deconfinement can be understood as a center vortex percolation 
transition \cite{Chernodub:1998vk,Engelhardt:1999fd}.
\end{enumerate}

	Other gauges work as well; general conditions 
a suitable gauge has to fulfill were formulated in~\cite{Faber:1999gu}.
Here we would just like to briefly summarize an interesting insight into
the meaning of MCG fixing and center projection, due 
to~\cite{Engelhardt:1999xw,Faber:2001hq}.

%
\section{Best-fit Interpretation of MCG}
	Running a MC simulation, one can ask for the 
pure gauge configuration closest, in configuration space, 
to a given lattice gauge field:
\begin{equation}\label{Landau}
U_\mu(x)\quad \sim \quad 
g(x) g^\dagger(x+\hat{\mu})\equiv U_\mu^{(0)}(x).
\end{equation}
It is easy to show that finding the optimal $g(x)$ 
is equivalent to the problem of fixing to the \textit{Landau gauge}.

	Let us now allow for {$Z_2$} dislocations in the gauge 
transformation, i.e.\ fit the lattice configuration
by a thin center vortex configuration: 
\begin{equation}
U_\mu^{vor}(x)\equiv 
g(x) Z_\mu(x)g^\dagger(x+\hat{\mu}), 
\qquad Z_\mu(x)=\pm 1
\end{equation}
{$U_\mu^{vor}(x)$} becomes a continuous pure gauge in the 
{adjoint representation}, blind to the {$Z_\mu(x)$} factor.
One can make the fit in two steps:

	1.\ Determine {$g(x)$} up to a {$Z_2$} 
transformation by minimizing the square distance {$d^2_A$} between
{$U_\mu$} and {$U_\mu^{vor}$} in the adjoint representation,
%
%
which is easily seen to be equivalent to fixing to direct MCG.

	2.\ Find {$Z_\mu(x)$} by minimizing:
\begin{equation}
	\mbox{Tr}
\Bigl\{\left[U_{\mu}(x)-g(x) Z_\mu(x) g^\dagger(x+\hat{\mu})\right]
\left[U_{\mu}^\dagger(x)-g(x+\hat{\mu}) Z_\mu(x) g^\dagger(x)\right]\Bigr\},
\end{equation}
which requires the center projection prescription, Eq.\ (\ref{projection}).
%
%

	Summarizing, the procedure of direct MCG fixing + center projection
represents the best fit of a lattice configuration by a set of thin center
vortices.

%
\section{{Why Does MCG Sometimes Fail to Find Vortices?}}
	MCG fixing suffers from the Gribov copy problem. The 
iterative gauge-fixing procedure converges to a local maximum 
which will be slightly different for every gauge copy of a given 
lattice configuration.

	At the first sight, the problem seemed quite innocuous: We 
observed in~\cite{DelDebbio:1998uu}
that vortex locations in random copies of a given configuration
were strongly correlated. However, the successes of the approach 
were seriously questioned. Bornyakov et al.~\cite{Bornyakov:2000ig} 
showed that using the method of simulated annealing instead of our
usual (over-)relaxation, one could find better (local) MCG maxima, 
but the center-projected string tension was only about 2/3 of the full 
one.

	The best-fit interpretation of the previous section provides 
us with a clue to the origin of this problem.
It is clear that $U_\mu^{vor}(x)$ is a bad
fit to $U_\mu(x)$ at links belonging to thin vortices (i.e.\ to the
P-plaquettes formed from $Z_\mu(x)$). We recall that a 
plaquette $p$ is a P-plaquette iff $Z(p)=-1$ (where $Z(C)$ denotes the
product of $Z_\mu(x)$ around the contour~$C$) and that  P-plaquettes belong
to P-vortices. Let us write the gauge transformed configuration as 
\begin{equation}
{}^g U_\mu(x) = Z_\mu(x)\;e^{iA_\mu(x)},
\qquad \mbox{Tr}\;e^{iA_\mu(x)}\ge0.
\end{equation}
At large $\beta$ values
$\textstyle{\frac{1}{2}}\mbox{Tr}[U_P]=1-O(1/\beta)$,
and equals to
\begin{equation}
\left(Z_P\right)
\textstyle{\frac{1}{2}}
\mbox{Tr}\displaystyle{\prod_P} e^{iA_\mu(x)} 
= {}_{~\atop\mbox{{\small\it on P-plaquettes\ }}}
{(-1)}\times\textstyle{\frac{1}{2}}
\mbox{Tr}\displaystyle{\prod_P} e^{iA_\mu(x)}.
\end{equation}
The last equation implies that at least at one link belonging to the 
P-plaquette $A_\mu(x)$ cannot be small, therefore ${}^g U_\mu(x)$ 
must strongly deviate 
from the center element. 

	The above argument shows that the quest for the global maximum 
may not always be the best strategy; one should rather try to exclude 
contributions from P-plaquettes where the fit is inevitably bad~
\cite{Faber:2001hq}, or modify the gauge fixing procedure to soften 
the fit at vortex cores.

%
%
\section{A Cure for the Disease: Direct Laplacian Center Gauge}
	We have recently proposed to overcome the Gribov problem
using the \textit{direct Laplacian center gauge}~\cite{Faber:2001zs}. 
The proposal was to a large extent inspired by the Laplacian 
Landau~\cite{Vink:1992ys}, Laplacian abelian~\cite{vanderSijs:1997hi}, 
and Laplacian center~\cite{Alexandrou:1999iy} gauges. 
The idea is the following: 

	To find the ``best fit'' to a lattice configuration by a 
thin center vortex configuration one looks for a matrix $M(x)$ 
maximizing the expression:
\begin{equation}
{\cal{R}}_M= \displaystyle{\sum_{x,\mu}} 
\mbox{Tr}\left[M^T(x) U_{A\mu}(x) M(x+\hat{\mu})\right],
\end{equation}
with a constraint that $M(x)$ should be an SO(3) matrix at
any site $x$:
\begin{equation}
M^T(x)\cdot M(x)= {\mathbf{1}},\qquad\det M(x)=1.
\end{equation}
We soften the orthogonality constraint by demanding it
only \textit{``on average''}:
\begin{equation}
\langle M^T\cdot M\;\rangle\equiv
\displaystyle{\frac{1}{\cal V}\sum_x} M^T(x)\cdot M(x)= \mathbf{1}.
\end{equation}

	It is convenient to write the columns of $M(x)$ as a set of 3-vectors:
$f^{b}_a(x) = M_{ab}(x)$.
The optimal $M(x)$, maximizing {${\cal{R}}_M$} with the constraint, 
is determined by the \textit{three} lowest 
eigenvectors $f^b_a(x)$:
\begin{equation}
{\cal D}_{ij}(x,y) f^a_{j}(y) = \lambda_a f^a_i(x)
\end{equation}
of the covariant adjoint Laplacian operator  
${\cal D}_{ij}(x,y)$:
\begin{equation}
{\cal{D}}_{ij}(x,y) = \sum_{\mu}
\Bigl(2\delta_{xy}\delta_{ij} 
- \left[U_{A\mu}(x)\right]_{ij}\delta_{y,x+\hat\mu} 
- \left[U_{A\mu}(x-\hat\mu)\right]_{ji}\delta_{y,x-\hat\mu}\Bigr) .
\end{equation}
	The resulting real matrix field $M(x)$ has further to be
mapped onto an SO(3)-valued field $g_{A}(x)$. A \textit{naive map}
(which could also be called \textit{Laplacian adjoint Landau gauge})
amounts to choosing $g_{A}(x)$ closest to $M(x)$.
Such a map is well known in matrix theory and is called 
\textit{polar decomposition}.

\begin{figure}[t!]
\centering
\includegraphics[width=0.5\textwidth]{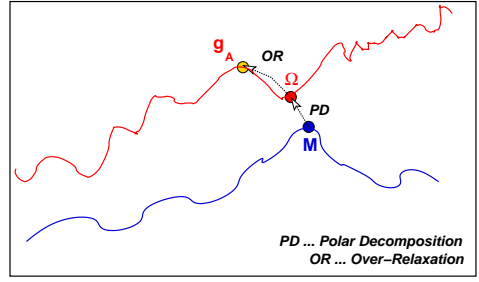} 
\caption{{Scheme of the Laplacian mapping of {$M(x)$} to
a nearby SO(3) matrix field {$g_A(x)$}.}}
\label{scheme}
\end{figure}

	A better procedure, in our opinion, is the
\textit{Laplacian map}, that leads to 
\textit{direct Laplacian center gauge}. We try to locate $g_A(x)$ as 
close to $M(x)$ \textit{local} maximum of the MCG (constrained)
maximization problem. To achieve this, we first make the naive map
(polar decomposition),
then use the usual quenched maximization (overrelaxation) to relax to
the nearest (or at least nearby) maximum of the MCG fixing condition.
This procedure is illustrated schematically in Fig.~\ref{scheme}.

	To test the new procedure, we have recalculated the
vortex observables introduced in our previous work 
(cf.\ Refs.\ \cite{DelDebbio:1996mh,DelDebbio:1998uu}), 
with P-vortices located via center projection after fixing the lattice to 
the new direct Laplacian center gauge. The results are summarized on 
the following page.

\begin{figure}[p!]
\centering
\begin{minipage}[t]{0.48\textwidth}
\centering
~

\includegraphics[width=\textwidth]{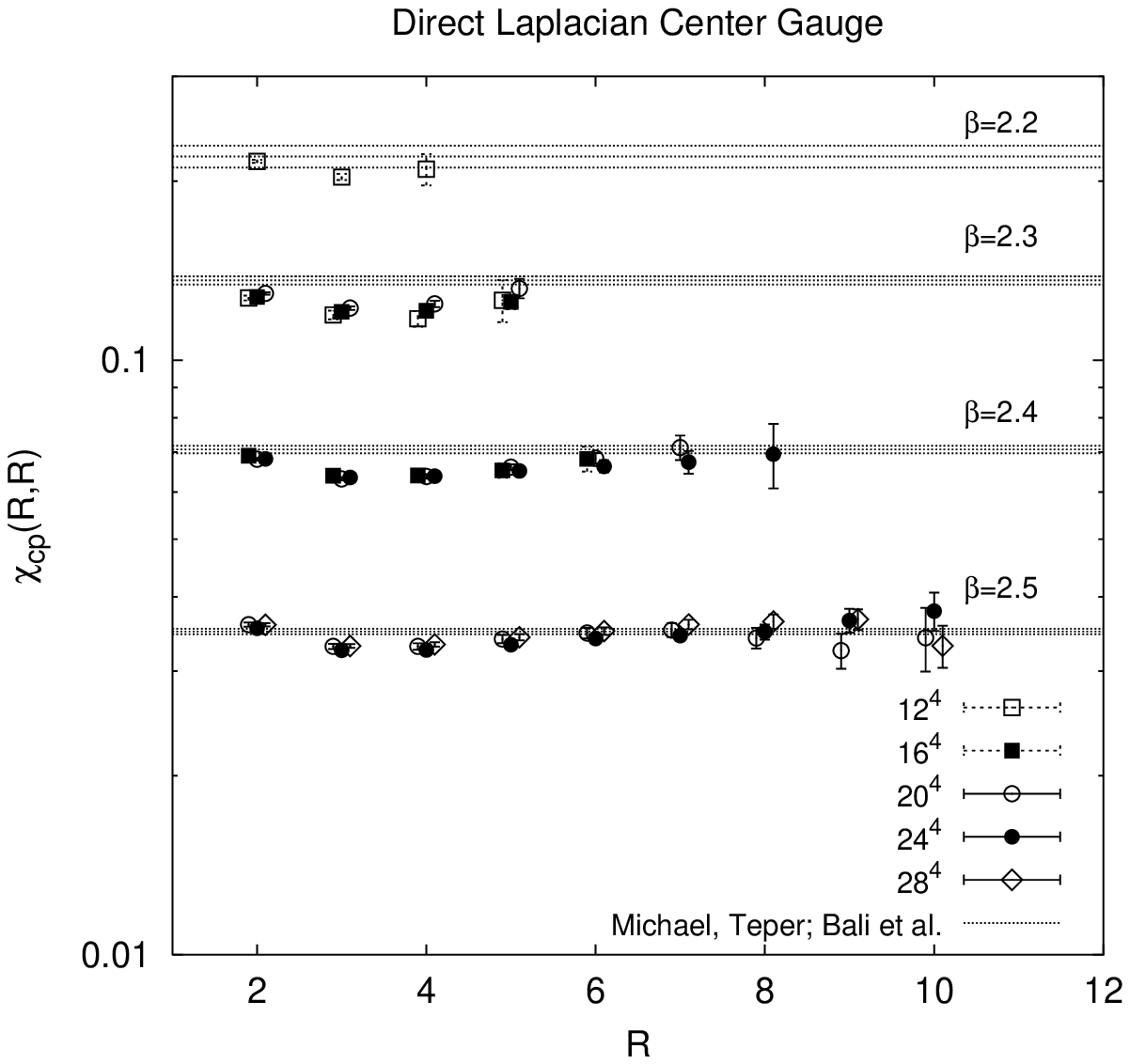}\\
\centerline{\scriptsize (a)}
\vspace*{0.1cm}
\includegraphics[width=\textwidth]{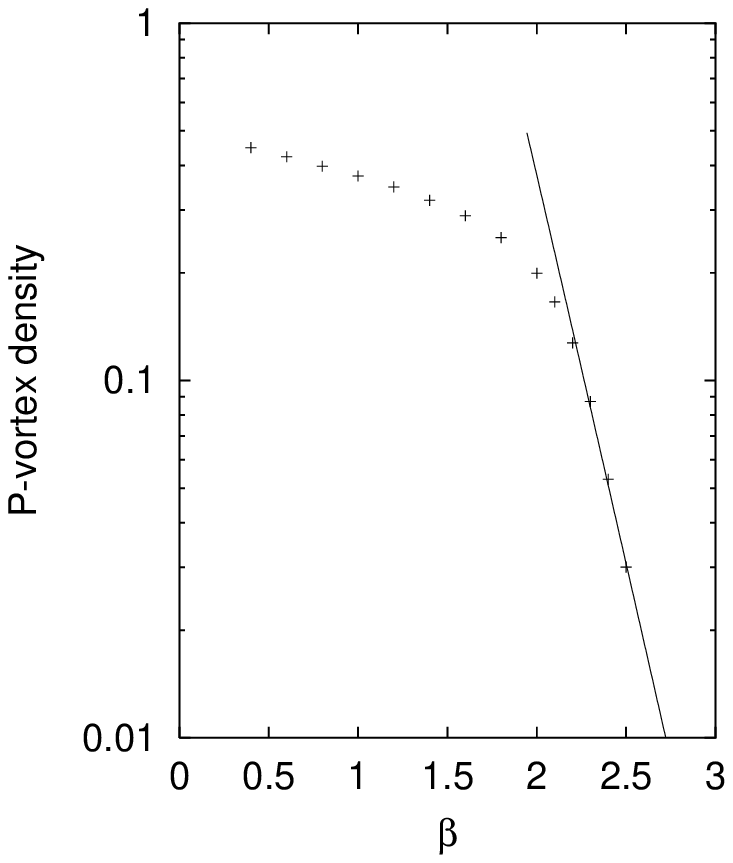}\\
\centerline{\scriptsize (c)}
\end{minipage}
\begin{minipage}[t]{0.01\textwidth}
\centering
~
\end{minipage}
\begin{minipage}[t]{0.48\textwidth}
\centering
~ 

\includegraphics[width=\textwidth]{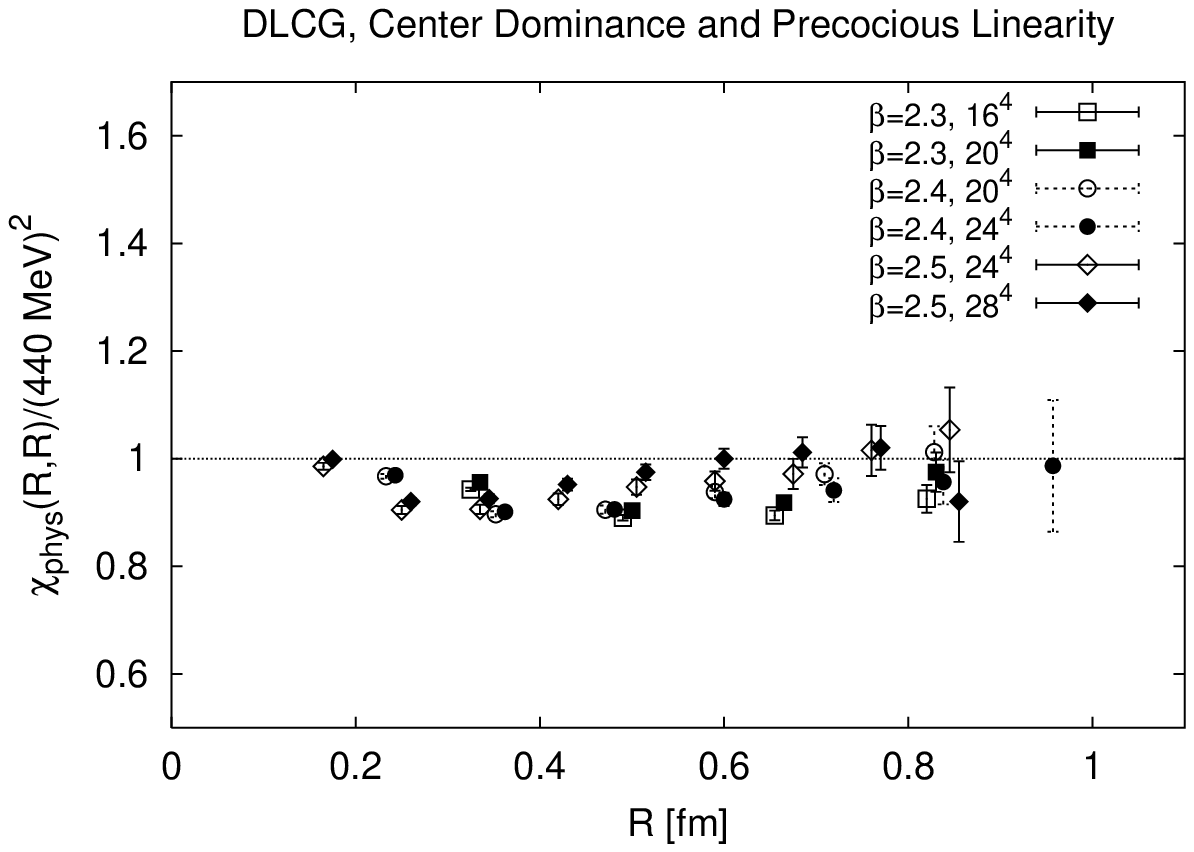}\\
\centerline{\scriptsize (b)}

\vspace*{0.1cm}
\includegraphics[width=\textwidth]{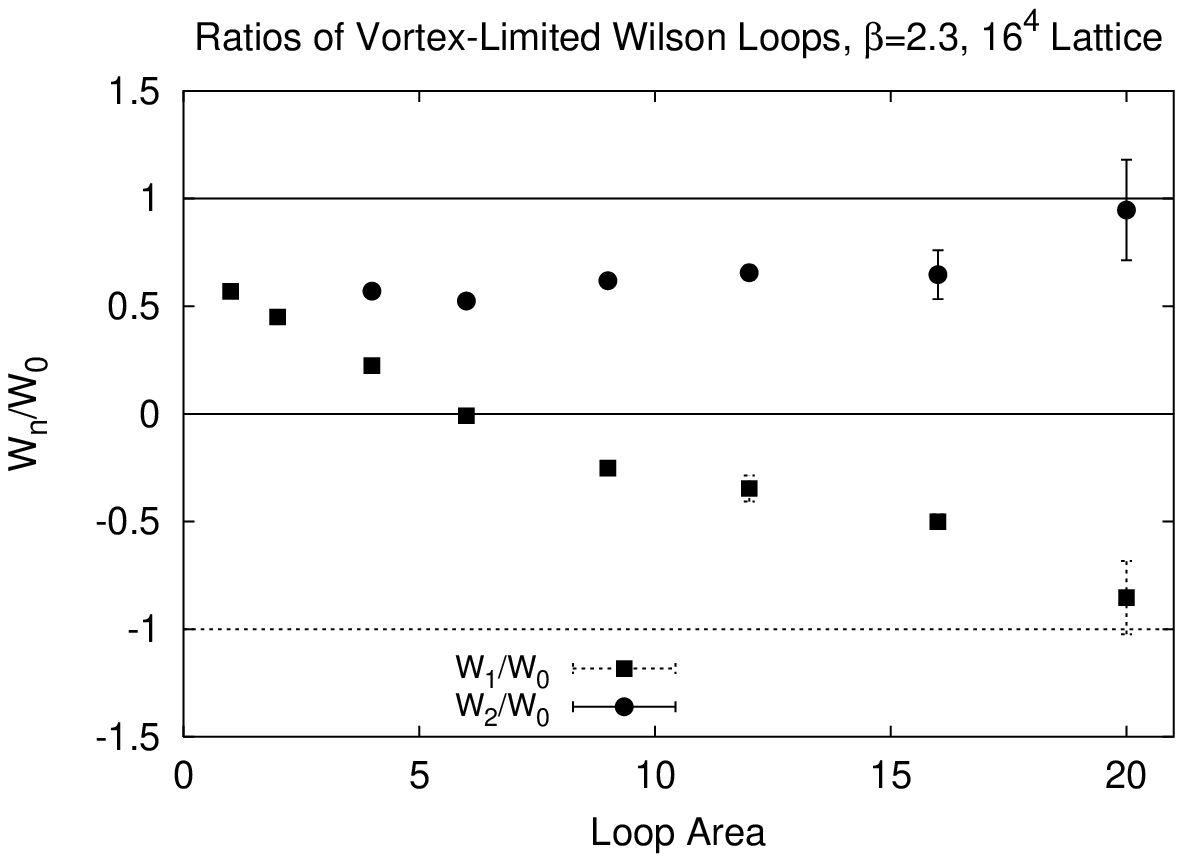}\\
\centerline{\scriptsize (d)}

\vspace*{0.1cm}
\includegraphics[width=\textwidth]{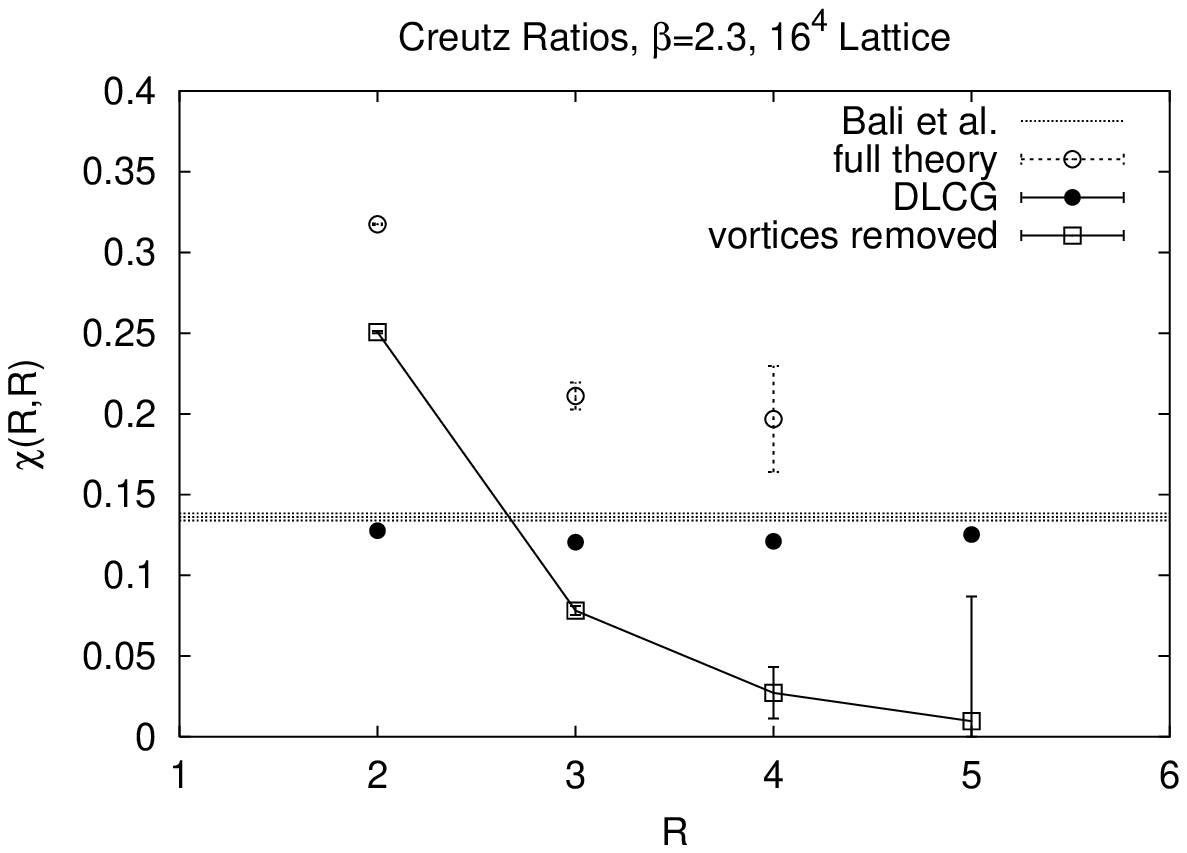}\\
\centerline{\scriptsize (e)}
\end{minipage}
\caption{Results from direct Laplacian center gauge:\newline
(a) Combined data, at {$\beta=2.2-2.5$}, for center-projected
Creutz ratios obtained after direct Laplacian center gauge fixing.
Horizontal bands indicate the asymptotic string tensions on the
unprojected lattice, with the corresponding errorbars, taken
from \protect\cite{Michael:fu}.\newline
(b) The ratio of projected Creutz ratios to the full asymptotic
string tension, as a function of loop extension in fermis.  The
data is taken from {$\chi_{cp}(R,R)$} at a variety of couplings and
lattice sizes.\newline
(c) Evidence of asymptotic scaling of the P-vortex surface
density.  The solid line is the asymptotic freedom prediction 
with {$\sqrt{\rho / 6\Lambda^2} = 50$}.\newline
(d) Ratio of one- and two-vortex to zero-vortex Wilson loops
{$W_{1,2}(C)/W_0(C)$} vs.\ loop area, at {$\beta=2.3$} 
on a {$16^4$} lattice.\newline
(e) Creutz ratios on the modified
lattice, with vortices removed, at {$\beta=2.3$}.}
\label{DLCG}
\end{figure}

	The quantities of the most immediate interest 
are the center-projected Creutz ratios. Our 
data for the range of couplings $\beta=2.2-2.5$ is displayed on a
logarithmic plot in Fig.\ \ref{DLCG}a.  In general
$\chi_{cp}(R,R)$ deviates from the full asymptotic string tension by
less than 10\%. 

	As another way of displaying both center dominance and
precocious linearity, we show, in Fig.\ \ref{DLCG}b, the ratio
\begin{equation}  
{\chi_{phys}(R,R)/\sigma_{phys}}          
= {\chi_{cp}(R,R)/\sigma_{Lat}(\beta)}
\end{equation}
as a function of the distance in physical units
$R_{phys} = R a(\beta)$  for all $\chi_{cp}(R,R)$ data points
taken in the range of couplings $\beta=2.3-2.5$.  Again we see that the
center-projected Creutz ratios and asymptotic string tension are in
good agreement (deviation $< 10\%$), and there is very little variation
in the Creutz ratios with distance. We should probably stress in this
context the significance of precocious linearity: it implies that 
center-projected degrees of freedom have isolated the long-range physics,
and are not mixed up with ultraviolet fluctuations.

	Other encouraging results from MCG are recovered in the new gauge 
as well. Figure \ref{DLCG}c shows the P-vortex density vs.\ $\beta$ in a 
logarithmic plot. The density scales according to the asymptotic freedom
formula with the slope corresponding to a quantity that behaves like a
surface density. The slope for pointlike objects (like instantons), or 
linelike objects (like monopoles) would be quite different.

	Figure \ref{DLCG}d presents the data on vortex-limited Wilson loops.
One can clearly see the expected trend, see Eq.\ (\ref{Wn/W0}), 
for large enough loops. Figure \ref{DLCG}e shows that removal 
of center vortices causes the asymptotic string tension to vanish. 

%
\section{How Does DLCG Differ from Laplacian Center Gauge?}
	The first step of direct Laplacian center gauge fixing 
is similar to the Laplacian center gauge proposed by de Forcrand and
collaborators~\cite{Alexandrou:1999iy}. Instead of using the three lowest 
eigenvectors of the covariant adjoint Laplacian operator and 
the naive map (or polar decomposition, see above), de Forcrand et al.\ 
build on the \textit{two} lowest eigenvectors only. The gauge is fixed by
$g(x)$ that
\begin{enumerate}
\item
	makes the lowest lying eigenvector to point in the 
third color direction (U(1) invariance still remains), and
\item
	rotates the second lowest eigenvector into (say) 
the first color direction.
\end{enumerate}
There is an ambiguity in the procedure when the first and
second vectors are collinear, and such ambiguities should define
positions of center vortices.

	LCG has its virtues and vices. It is unique (apart from eventual
true Gribov copies) and shows center dominance after center projection.
On the other hand, center dominance is seen only for very large distances,
and there is not a good separation between confinement and short-range
physics: there is no precocious linearity, there are too many 
vortices, vortex density does not scale. Moreover, identification 
of vortices via gauge fixing ambiguities fails for simplest configurations 
(like a pair of thin vortices put on the lattice by hand~\cite{Faber:1999gu}), 
and is practically impossible in Monte-Carlo generated configurations. 
Center projection is necessary.

	To improve on these problems, Langfeld et al.~\cite{Langfeld:2001nz}
proposed to follow the LCG procedure of de Forcrand et al.\ by 
(over-)relaxation to MCG. This, in analogy with DLCG, could be called
\textit{indirect Laplacian center gauge}.\footnote{LCG involves first
fixing to Laplacian abelian gauge, then further reducing the residual
symmetry from U(1) to $Z_2$, in which it is reminiscent of indirect
maximal center gauge of Ref.~\cite{DelDebbio:1996mh}.}

\begin{figure}[t!]
\begin{minipage}[t]{0.48\textwidth}
\centering
\includegraphics[width=\textwidth]{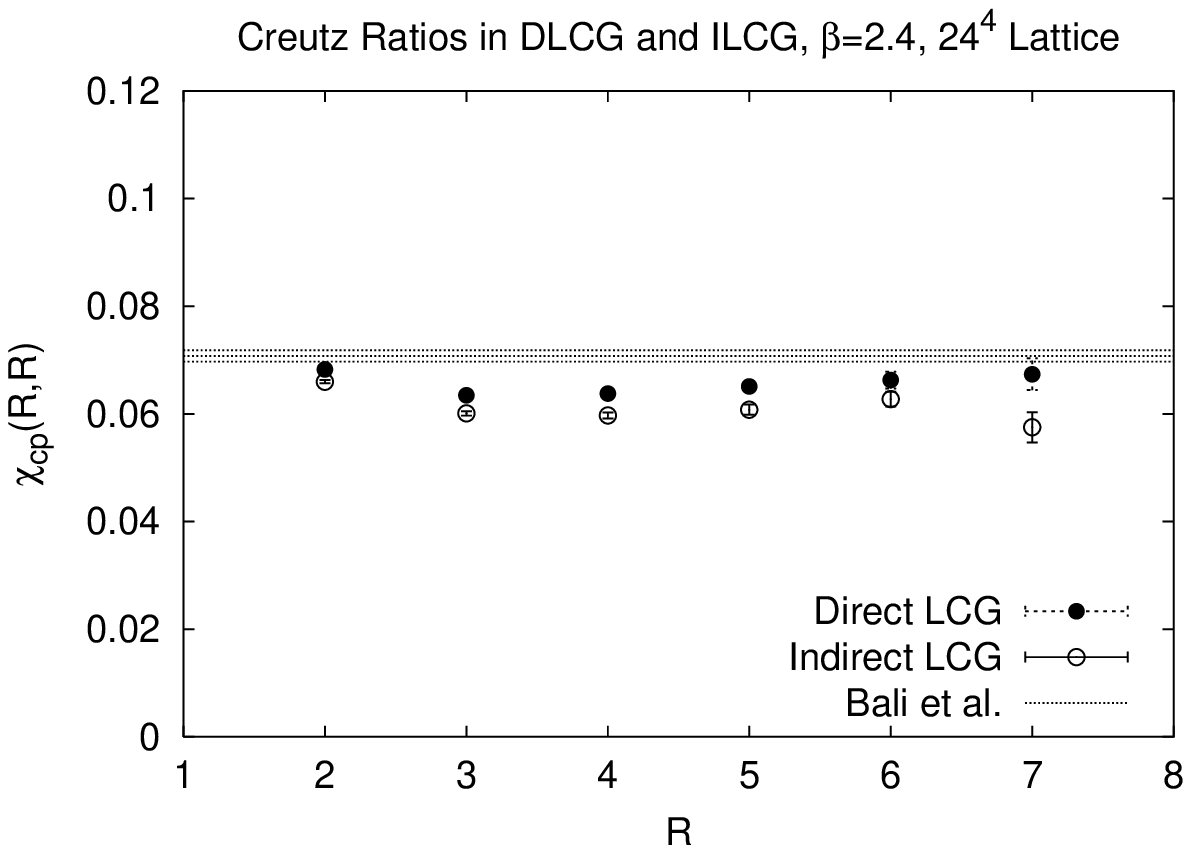}\\
\centerline{\scriptsize (a)}
\end{minipage}
\begin{minipage}[t]{0.01\textwidth}
\centering
~
\end{minipage}
\begin{minipage}[t]{0.48\textwidth}
\centering
\includegraphics[width=\textwidth]{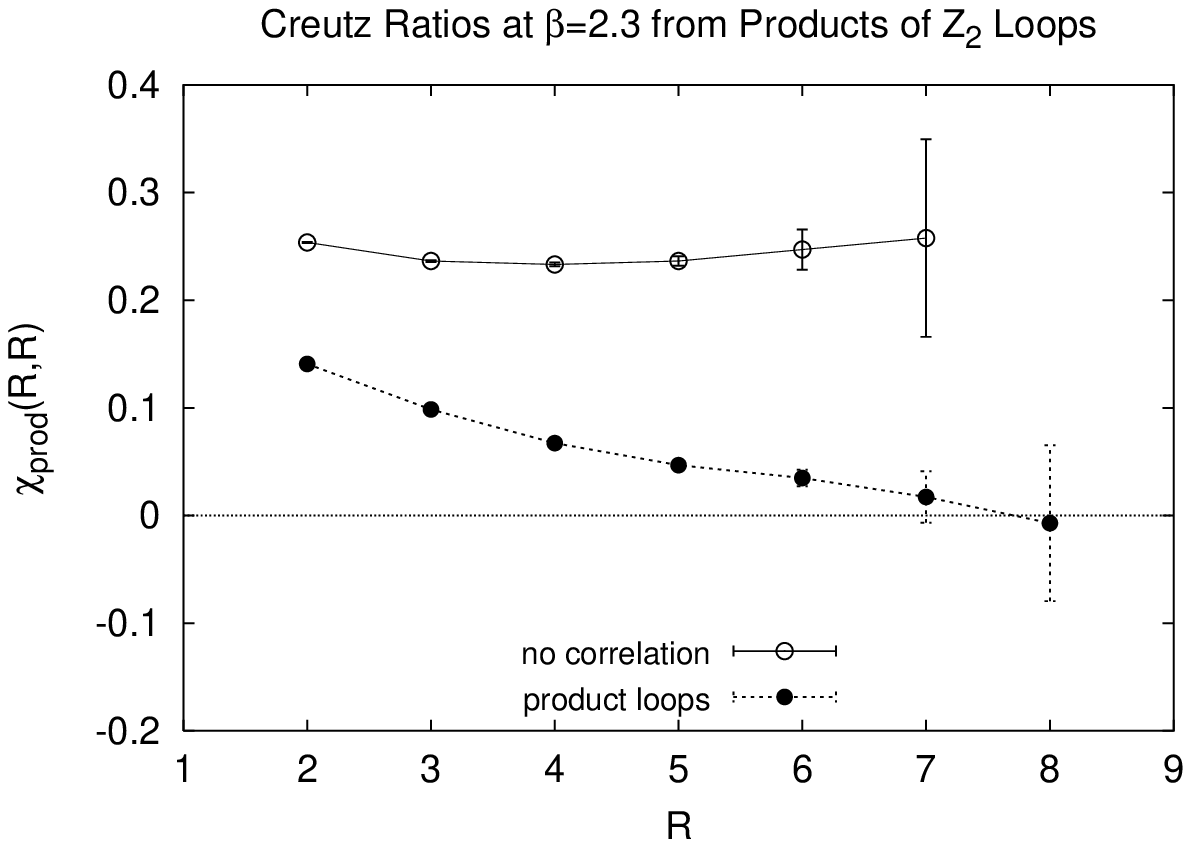}\\
\centerline{\scriptsize (b)}
%
\end{minipage}
\caption{Comparison of the direct and indirect Laplacian center gauges:
\newline
(a) Center projected Creutz ratios. (The asymptotic 
string tension is shown by the horizontal band.)\newline
(b) Creutz ratios $\chi_{prod}(R,R)$ calculated from
``product Wilson loops'', Eq.\ (\protect\ref{productloops}).}
\label{comparison}
\end{figure}

	The question is whether results from DLCG and ILCG differ 
considerably, and whether there is any correlation between
vortex locations in those two gauges. 
Figure~\ref{comparison}a shows the projected
Creutz ratios at $\beta=2.4$ in ILCG; for comparison we also display 
the corresponding data from DLCG. It seems that the center dominance 
properties are somewhat better in DLCG than in ILCG, though the difference is 
not great. The reason for this is quite easy to explain: both procedures seem 
to locate the same physical vortices. 

	The simplest way to test
the last statement is the following: For a given lattice $\{U_\mu(x)\}$
let $\{Z'_\mu(x)\}$ be the lattice obtained by center projection in DLCG,
while $\{Z''_\mu(x)\}$ be the corresponding lattice in ILCG. Denoting
by $Z'(C)$ and $Z''(C)$ the Wilson loops in these two projected lattices,
we construct the ``product'' loops
\begin{equation}\label{productloops}
Z_{prod}(C)=\langle Z'(C)\;Z''(C)\rangle
\end{equation}
and from their expectation values the corresponding Creutz ratios
$\chi_{prod}(I,J)$. The expectation is that if the two projected lattices were
perfectly correlated
\begin{equation}
\chi_{prod}(R,R)=0 \qquad \mbox{(perfect correlation),}
\end{equation}
whereas in case of zero correlation
\begin{equation} 
\chi_{prod}(R,R)=\chi'_{cp}(R,R)+ \chi''_{cp}(R,R) 
\qquad \mbox{(no correlation).}
\end{equation}

	It is evident from Fig.~\ref{comparison}b that center-projected loops
in both gauges are not well correlated at short distances, but become 
correlated at large distances. The interpretation, we believe, 
is straightforward: The
P-vortices in each projected lattice do not coincide, but in most cases
are located within the same (thick) center vortices on an unprojected lattice.
This accounts for the strong correlation on large distance scales, larger than
the typical size of vortex cores. Similar correlations exist also
with projected lattices in MCG (with gauge fixing via overrelaxation).

%
%
\section{Summary}
\begin{enumerate}
\item
	Center dominance exists in various gauges. The maximal center 
gauge has an appealing ``best-fit'' interpretation, but 
the successes of the approach have been overshadowed by the
problem of Gribov copies. 
\item
	We have proposed a new gauge, 
direct Laplacian center gauge, that combines
fixing to adjoint Laplacian Landau gauge with the usual overrelaxation. 
The first step of the procedure is unique, in the second step no
strong gauge-copy dependence appears. This procedure can be 
interpreted as a ``best fit'' softened at vortex cores.
\item
	All features known from MCG are reproduced in direct LCG:
center dominance, precocious linearity, scaling of the vortex density,
etc. 
\item
	Similar results follow from center projection in Laplacian center 
gauge after overrelaxation (indirect LCG). The reason is that
vortex locations in projected lattices in direct and indirect 
LCG are quite strongly correlated.
\end{enumerate}
%

%

%

%

\end{document}